\documentstyle[12pt]{article}
\def\beq{\begin{equation}}
\def\eeq{\end{equation}}
\def\bea{\begin{eqnarray}}
\def\eea{\end{eqnarray}}
\def\bq{\begin{quote}}
\def\eq{\end{quote}}

\def\JP{{\it J.Phys.} }

\def\NP{{\it Nucl.Phys.} }
\def\PL{{\it Phys.Lett.} }
\def\PR{{\it Phys.Rev.} }

\def\ZP{{\it Z.Phys.} }

\def\gappeq{\mathrel{\rlap {\raise.5ex\hbox{$>$}}
{\lower.5ex\hbox{$\sim$}}}}

\def\lappeq{\mathrel{\rlap{\raise.5ex\hbox{$<$}}
{\lower.5ex\hbox{$\sim$}}}}

\begin{document}
\pagestyle{empty}

\vspace*{2cm}
\begin{center}
{\bf INTERCONNECTION PHENOMENA IN $W^+W^-$ and $t\bar t$ EVENTS}\\
\vspace*{2.5cm}
Valery A. KHOZE \\
\vspace*{0.8cm}
{\it INFN - Laboratori Nazionali di Frascati,}\\ 
{\it P.O. Box 13, I-00044 Frascati (Roma) Italy}\\
{\it and}\\
{\it Theoretical Physics Division, CERN,\\ CH - 1211 Geneva 23,
Switzerland} \\
\vspace*{6.5cm}
{\bf Abstract}
\end{center}

I will attempt to survey some selected physics issues on QCD
interconnection
 phenomena in the processes $e^+e^-\rightarrow W^+W^-\rightarrow$ 4 jets
and
$e^+e^-\rightarrow t\bar t \rightarrow b W^+\bar b W^-$. 
Possible consequences for LEP2 and future linear $e^+e^-$ colliders are 
briefly discussed.

\vfill\eject

\noindent
{\bf 1~~~ Introduction}

It is widely believed that in particle physics \lq\lq tomorrow belongs" to 
the detailed studies of heavy unstable objects.  Firstly, we anticipate
the 
exciting discoveries of new heavy particles (Higgs boson(s), SUSY
particles, 
$W^{\prime},Z^{\prime}$,...) at increasingly higher energies.  Secondly,
for the 
precision tests of the Standard Model one needs the high accuracy
determination 
of the parameters of the $W$ boson and of the top quark, primarily their
masses. 

Let us briefly address the latter point.  These years we have witnessed
some
important developments in precision electroweak tests. One can consider as
an 
impressive success of the Standard 
Model the fact that the top mass $m_t$ predicted from the electroweak data 
agrees within the stated errors with the  direct Tevatron result
\cite{one}
\beq
m_t = 173, 8 \pm 5,2~{\rm GeV}
\label{un}
\eeq

There has also been further progress with the determination of the $W$
boson mass
$m_W$ at the Tevatron \cite{two}. Preliminary results on the Tevatron
average is 
\beq
m_W = 80.41  \pm 0.09~{\rm GeV}
\label{deux}
\eeq

Meanwhile, at LEP2 the combined statistical uncertainty on $m_W$ has
reached a
level of about 70 MeV \cite{three}-\cite{six}.

The precise measurements of $m_W$ and $m_t$ is a priority of present and
future
experimental studies. These will allow to fully exploit the remarkable
accuracy
of exploring the $Z^0$ physics and other precision electroweak
measurements. One
may hope to pin down the Higgs mass or/and to look for evidence for
physics
beyond the Standard Model.

What are the prospects of the experimental studies?

Run~II at the Tevatron and LEP2 are aiming for an uncertainty on $m_W$ of
about
35-40 MeV, see Refs. \cite{two},\cite{five}. An upgrade of the Tevatron,
beyond
Run~II, and the LHC may allow a pricision on $m_W$ of about 15 MeV, see,
e.g.,
\cite{seven}. It seems reasonable to expect that future experiments at the
Tevatron and the LHC will increase an accuracy of $m_t$ measurements up to
1-2
GeV.

A unique precise determination of $m_t$ (with an accuracy of a few 
hundred MeV) will be 
one of the most attractive physics topics at future linear $e^+e^-$ 
and muon colliders [8]-[10]. 

An obvious requirement for success of these precise studies is that the
accuracy 
of the theoretical predictions should match or better exceed the
experimental 
errors.  This requires a detailed understanding of production and decay 
mechanisms and, in particular, of the effects arising from the large
width, 
$\Gamma \sim O$ (1 GeV).  Recall that in production processes of heavy
unstable 
particles it is natural to separate the production stage from the decay 
processes.  In general these stages are not independent and may be 
interconnected by radiative interference effects.  Particle(s) (e.g.\
gluon(s) 
and/or photon(s)) could be produced at one stage and absorbed at another;
we 
speak of virtual interference.  Real interference will occur as well since
the 
same real particle can be emitted from the different stages of the
process. 

Many observations rely on a clear understanding of the role of these 
interference effects.  Indeed there is a long list of examples where a
detailed 
knowledge of interferences can be important for the interpretation of 
experimental data (see [11]-[13] and references therein).

In this talk I concentrate mainly on the QCD interconnection phenomena
that may 
occur when two unstable particles ($W$ bosons, top quarks) decay  
close to each other.  The word \lq interconnection' is here 
introduced to cover those aspects of final-state particle production that
are 
not dictated by the separate decays of unstable objects, but can only be 
understood in terms of the joint action of the two.
Such a cross-talk between heavy unstable particles could occur because
they decay
at short distances of order $1/\Gamma \sim$ 0.1 fm, and their decay
products
hadronize close to each other in space and time at the typical hadronic
scale of
$\sim$ 1 fm.

\vspace*{1cm}
\noindent
{\bf 2~~~ QCD Interconnection in Hadronic $W^+W^-$ Events}

The accurate determination of the W-boson mass is one of the main
objectives of LEP2.  However, the systematic
uncertainties  due to hadronic final-state interactions and QCD
interferences  between the W decay products may induce substantial
ambiguities,  for reviews see Refs. [14],[15]. 

The cross-talk between the $W^{\pm}$ decay products undermines the
traditional meaning of a W mass in the process
\begin{equation}
e^+e^- \to W^+ W^- \to q_1 \bar q_2 q_3 \bar q_4 ~,
\label{trois}
\end{equation}
called the (4q) mode.
It is not even in principle possible to subdivide the final-state
hadrons into two groups, one of which corresponds to the
$W^+ \to q_1 \bar q_2$ decay and the other to the
$W^- \to q_3 \bar q_4$ decay: the identities of individual
$W^{\pm}$ decay products are not well-defined any more.
If the $W$-boson lifetime could be considered as very short, $1/\Gamma_W 
\rightarrow 0$, both the $q_1\bar{q}_2$ and $q_3\bar{q}_4$ pairs appear
almost 
instantaneously, and they radiate coherently, as though produced at the
same 
vertex.  In the other extreme, $\Gamma_W \rightarrow 0$, the
$q_1\bar{q}_2$ and 
$q_3\bar{q}_4$ pairs appear at very different times $t_1,t_2$ after the
$W^+W^-$ 
production, 
\begin{equation} 
\tau_p \sim \frac{1}{m_W} \ll \Delta t = |t_1 - t_2| \sim
\frac{1}{\Gamma_W} . 
\label{quatre}
\end{equation} 
The two dipoles therefore radiate gluons and produce hadrons according to
the 
no-reconnection scenario. 

The crucial point is the proper choice of the scale the $W$ width should
be 
compared with.  That scale is set by the energies of primary emissions,
real or 
virtual, see Ref. [11] and references therein.   Let us clarify this
supposing, for simplicity, that we are  in the $W^+W^-$ threshold region.
The
relative phases of radiation accompanying  two $W$ decays are then given
by the
quantity 
\begin{equation} 
\omega_i \Delta t \sim \frac{\omega_i}{\Gamma_W} . 
\label{cinq}
\end{equation} 
When $\omega_i/\Gamma_W \gg 1$ the phases fluctuate wildly and the
interference 
terms vanish.   The argumentation remains valid for energies above the
$W^+W^-$ 
threshold as well. 

An instructive Gedanken experiment to highlight the filtering role of
$\Gamma$ 
can be obtained [13] by comparing the emission of photons in the eV to MeV
range
for  the two processes 
\begin{equation} 
\gamma\gamma \rightarrow W^+W^- \rightarrow \mu^+ \nu_{\mu} \mu^- 
\bar{\nu}_{\mu} , 
\label{six}
\end{equation} 

\begin{equation} 
\gamma\gamma \rightarrow K^+K^- \rightarrow \mu^+ \nu_{\mu} \mu^- 
\bar{\nu}_{\mu} , 
\label{sept}
\end{equation}       
near threshold, in the extreme kinematical configuration where the $\mu^+$
is 
collinear with the $\mu^-$.  For the first process, $\omega \ll \Gamma_W$,
and 
one expects hardly any radiation at all, because of the complete screening
of 
the two oppositely charged muons.  For the second process, $\omega \gg 
\Gamma_K$, the parent particles have long lifetimes and the $\mu^+$ and
$\mu^-$ 
appear at very different times.  The photon wavelength is very small
compared 
with the size of the $\mu^+\mu^-$ dipole and, therefore, the $\mu^+$ and
$\mu^-$ 
radiate photons independently, with no interference. 

Suppression of the interference in the case of radiation with $\omega_i
\gg 
\Gamma_W$ can be demonstrated also in a more formal way, see e.g., [11]. 

Note that the limiting case when $\tau_{\rm dec}\sim {1\over\Gamma_W} \sim
\tau_p
\sim {1\over m_W}$ represents an example of the so-called instantaneous
reconnection scenario, where the alternative colour singlets are
immediately
formed and allowed to radiate perturbative gluons with an energy up to $O
(m_W)$, see for details Refs. [12],[16].

The strong-interaction dynamics induces a variety of interconnection
effects between the hadronic decays of different W's, such as:
\begin{enumerate}
\item
Quantum short-distance effects due to exchanges of perturbative gluons
between the two initial $q\bar q$ systems.
\item
Final-state radiative gluon interferences on the stage of
parton-shower development.
\item
Long-distance effects in the parton-to-hadron transition phase
caused by a large overlap between the products of the two decays
(non-perturbative rearrangement/reconnection).
\item
Bose--Einstein (BE) correlations betwen identical bosons
(in practice, pions).
\end{enumerate}

The possibility of colour rearrangement in the
process (\ref{trois}) was first considered in Ref. [16].
The r\^ole of QCD interconnection in hadronic WW decays in the
framework of the W mass measurement was first
discussed in Ref. [12].
This challenging topic has been quite intensively studied theoretically
since then. We do not attempt to cover the consequences of
BE effects here, see Refs. [17].

It is necessary to emphasize that there is no question of whether
interconnection between the W's exists or not; it is certainly there
even in the QED context. Thus, the final-state QED interconnection induces
a
sizeable mass  shift, $(\delta m_W)_{\rm QED} \sim O(\alpha_{em} \pi
\Gamma_W)
\sim 50$~MeV, in $e^+e^- \to 4$~fermions in the threshold region
\cite{eighteen},\cite{nineteen}. However, at energies above 170~GeV,
$(\delta m_W)_{\rm QED} \sim O(\alpha_{em} \Gamma_W / \pi)$,
and cannot exceed a few MeV \cite{nineteen}-\cite{twentyone}.
Another well-known precedent is ${\rm J}/\psi$ production in
$B$ decay: the $c\bar c  \to {\rm J}/\psi$ transition requires
a cross-talk between the two original colour singlets,
$\bar c+s$ and $c + {\rm spectator}$.
The real challenge is to understand how large the ambiguities
for various observables can be. Evidently, it is not only the W
mass that can be affected by interconnection. Various event
characteristics in hadronic WW decay  could, in principle, show effects
even  an order of magnitude bigger than that in $m_W$.
On the other hand, in the inclusive cross section for process
(\ref{trois}), the effects of the QCD (and QED) cross-talk are
negligible \cite{twentytwo}:
\begin{equation}
\frac{\Delta \sigma_{WW}^{\rm intercon}}{\sigma_{WW}} \sim
\left( \frac{C_F \alpha_s(\Gamma_W)}{\pi} \right)^2
\frac{1}{N_C^2} \frac{\Gamma_W}{m_W} ~,
\label{huit}
\end{equation}
where $C_F = (N_C^2-1)/2N_C$, $N_C=3$ being the number of colours.

A precise measurement of the $e^+e^- \to W^+ W^-$ threshold
cross section (see  e.g., Ref. [14] for details) would provide an
interconnection-free method for measuring $m_W$. Unfortunately,
the combined total luminosity accumulated at LEP2 at
 $\sqrt{s} = 161$~GeV
is not sufficient to reach an interesting level of precision.
So the direct kinematic reconstruction of $m_W$ from the W
hadronic decays remains the only realistic method at current and
future energies of LEP2, see Refs. [4]-[6].

The potential significance of the cross-talk phenomena for the W mass
reconstruction at LEP2 obviously warrants a detailed
understanding of the size of the corresponding ambiguities. Note also
that QCD reconnection is of interest in its own right, since it may
provide us with a prospective laboratory for probing hadronization
dynamics in space and time.

The perturbative aspects of  QCD interconnection are, in principle,
well controllable. Since the corresponding W mass shift is expected
to be well within the uncertainties of the hadronization models (and
about on the same level as QED corrections) we only recall here an
estimate of Ref. [12],
\begin{equation}
(\delta m_W)_{\rm PT} \sim
\left( \frac{C_F \alpha_s(\Gamma_W)}{\pi} \right)^2
\frac{1}{N_C^2} \Gamma_W ~,
\label{neuf}
\end{equation}
which is of order of a few MeV. The perturbatively calculated
mass-shift (as well as other observables) is colour suppressed,
by two powers of $N_C$, which is typical for the gluon-mediated
interaction between the two colour-singlet objects.

In the non-perturbative stage, which is our main concern,
the colour-suppression situation varies between
scenarios. Here factors like $1/N_C^2$ may present,
as in the perturbative phase, but they are multiplied by
model-dependent coefficients, which are functions of the space--time
variables. These coefficients, in principle, could be
anything, even much larger than unity.

Since the space--time separation between the W$^+$ and W$^-$ decay
vertices is typically of order $\tau_{\rm dec} \sim 1/\Gamma_W$, only
rather soft
gluons (real or virtual) with an energy $\omega \lappeq \Gamma_W$ could
feel the
collective action of both the $q_1 \bar q_2$ and $q_3 \bar q_4$
antennae/dipole
systems, and thus participate in the cross-talk.
This explains the origin of the last factor in eq.~(\ref{neuf}).

Non-perturbative reconnection
can occur wherever the hadronization regions of the two W bosons
overlap. As was first emphasized in Ref. [12], the space--time
picture of the evolution of the final state plays an
essential r\^ole in understanding the size of the interconnection
effects at the hadronic level. At the moment,
the possible consequences of the hadronic cross-talk between the W's
can only be studied within the existing model-dependent
schemes of hadronization. These have done a very good job
in describing a vast amount of information on hadronic $Z^0$ decays,
so one may expect that (after appropriate modifications) they could
provide a reasonable estimate for the magnitude of interconnection-induced
effects, see Ref. [23] for a recent review.

The currently used algorithms for treating the non-perturbative cross-talk
 all assume a local interaction.
Reconnection-unrelated parameters are tuned to optimize the agreement
with $Z^0$ data. Some models  allow
reconnection also among the partons of a single $Z^0$, and then
consistency requires reconnection to be included in the above-mentioned
tuning stage.

Some essential phenomenological aspects appear to be common for different
interconnection models:
\begin{enumerate}
\item
The cross-talk dampens comparatively slowly with center-of-mass
energy, $\sqrt{s}$, over the range that can be tested by LEP2.
\item
Interconnection effects tend to be strongly dependent on the event
topology, and could induce azimuthal anisotropies in the particle flow
distributions.
\item
The low-momentum final particles ($p \lappeq 1$~GeV) are the main
mediators in the hadronic cross-talk, and they are most affected by it.
\item
Not far from the $WW$ threshold the invariant mass of an original
non-reconnected $q\bar q$ system is larger than that for a  reconnected
one. Therefore, most of the model predictions
show that the mean particle multiplicity in the (4q) mode,
$\langle N^{\rm (4q)} \rangle$, is lower than twice the mean
multiplicity of a hadronically decaying W in the mixed
hadronic--leptonic channel ((2q) mode),
$\langle N^{\rm (2q)} \rangle$,%
\begin{equation}
\frac{\langle N^{\rm (4q)} \rangle}{2\langle N^{\rm (2q)} \rangle}
 < 1 ~.
\label{dix}
\end{equation}
With increasing $\sqrt{s}$, the multiplicity in the purely hadronic
final state may start to rise \cite{twentyfour}.
However, at least within the models based on colour-confinement strings
 \cite{twentyfive}, the inequality (\ref{dix}) remains valid in the
whole range of LEP2 energies.
\item
All the models on the market (except of Ref. [26]) predict rather small
cross-talk effects. Thus, a conservative upper limit
on the $m_W$ shift seems to be something like around 50~MeV. Changes
in the standard global event characteristics are expected at the per cent
level. In marked difference with all other approaches, the colour-full
scenario of Ref. [26] allows much larger signals. Thus, the W mass
and the relative multiplicity shifts are predicted to be around 400~MeV
and 10\%, respectively. The strong claims of Ref. [26] have made the whole
subject of connectometry attractive for experimentalists.
\end{enumerate}

The word connectometry was introduced in Ref. [24] to cover various ways
to
detect inter\-connection-induced effects by measuring
characteristics of the WW final state. The first experimental
results on connectometry in the W$^+$W$^-$ events have already been
reported, and new experimental
information continues to pour out from LEP2, see  Refs. [4]-[6].
At the current level of statistics, there is no evidence for
interconnection effects from the standard distributions in hadronic
WW events. This agrees with the mainstream of model predictions,
which suggests rather small effects. However, it should be remembered
that a WW statistics larger by an order of magnitude is still
to come.

An important point to bear in mind is that the values -- even the
signs -- of shifts in various observables can depend strongly on the
hadronization scenario and on the choice of model parameters. Moreover,
results may be strongly sensitive to the adopted experimental strategy.

It would be extremely valuable to establish a model-independent
correlation between the shift in  $m_W$ and measurable quantities
in the final-state distributions. Unfortunately, so far studies do not
suggest any convincing correlation of such a type.
So one has to proceed within the framework of a certain QCD Monte
Carlo model. Thus, Ref. [24] attempted o quantify the expectations
 based on the string
hadronization model \cite{twentyfive} in terms of the distributions of
low-momentum hadrons. This idea was motivated by an observation
\cite{twelve}
that it is the soft particles that are  most sensitive to hadronic
cross-talk. Essential advantages of such an approach to connectometry
is that here the no-reconnection case can be well described, and that
there is  no (direct) dependence on the jet reconstruction method or
event selection strategy. Within string models, there are some
general qualitative predictions for the soft-particle spectra in the
$WW \to 4q$ events, $d n_{4q}^h / d p$, in the LEP2 energy range.
\begin{enumerate}
\item
Depopulation of the low-momentum hadrons, relative to the
no-reconnection scenario, due to the Lorentz boosts of the alternative
$q_1\bar q_4$ and $q_3\bar q_2$ dipoles/antennae.
\item
As in the case of the well-known standard string effect [25],[27],
such a depopulation should become more pronounced for heavier
hadrons (K, p, \ldots).
\item
A gradual reduction of the cross-talk with center-of-mass energy, since
the two outgoing W hadronic systems are more and more boosted apart.
\end{enumerate}

As it follows from Ref. [24], at $\sqrt{s}$ = 172 GeV within the realistic
hadronization scenarios, the depletion of low-momentum spectra, as
compared 
to the
no-reconnection case is
$\sim$ 2
\% for charged particles and
$\sim$ 5 \% for
$K+p$ ~~\footnote{On
experimental studies using tagged $K$ and $p$, see Ref. [28].}. At
$\sqrt{s}$ = 195  GeV, the effects drop down to about a half of what they
were at 172 GeV.

The studies in Ref. [24] do not encourage a too optimistic
 prognosis concerning the
prospects  of connectometry on the basis of low-momentum spectra, even
having the whole aimed-for statistics of LEP2. The best one can hope is
that the expected signal would be at the edge of observability. In such
a case one would need a lot of hard work (and good luck) in order to
detect the signal reliably. However, I would like to emphasize that it is
only experiment that could lead the way and may  cast light on the
challenging issues of the hadronic cross-talk.

I would also like to make it  clear that the nonobservation of
the reconnection effects on the low-momentum spectra, by no means,
indicates their nonexistence. Most likely it may just mean that the
``queen of observables" is still to be nominated.

Some other ideas may be useful, see Refs. [12],[23],[24]. For example, one
may
attempt an event-by-event reconstruction of the colour string/dipole
topology [30] (see also [29]).

Finally, let us recall that the $Z^0$
data provide an excellent experimental reference point,
thanks to LEP1. When the $Z^0$ results are used for calibration, the
actual model dependence of the low-momentum spectra proves to be
rather weak. Due to colour coherence in QCD cascades, the difference
in the evolution  scales corresponding to the $Z^0$ and the W
could cause only small changes (on the per cent level) at low momenta,
see Refs. [31],[32]. Effects due to the
difference in the primary quark flavour composition
also remain  on the per cent level for soft particles.  Such small
corrections could readily be accounted for, see Ref. [24] for details and
applications.

\vspace*{1cm}
\noindent
{\bf
 3~~~  Correlations of Particle Flow in Top Events}

One of the main objectives of a future linear $e^+e^-$ collider will be to 
determine the top mass $m_t$ with high accuracy.  Besides the traditional 
measurements of the $t\bar{t}$ excitation curve, several other approaches
are 
discussed, see, e.g., [10].  One method is to reconstruct the top
invariant 
mass event by event, another is to measure the top momentum 
distribution$^{[33]}$.  In either case, the QCD interconnection effects
could 
introduce the potentiality for a systematic bias in the top mass
determination.

It is not my intention to go here through all the details of the problem.
As a
specific topical example, following Ref. [13],
we consider the production and decay of a $t\bar{t}$ pair in the process 
\begin{equation}  
e^+e^- \rightarrow t\bar{t} \rightarrow bW^+\bar{b}W^- 
\label{onze}
\end{equation} 
and concentrate on the possible manifestations of the interconnection
effects in 
the distribution of the particle flow in the final state. 
For simplicity we assume that the $W$'s decay leptonically, so the colour
flow 
is generated only by the $t$ and $b$ quarks.  Further, we restrict
ourselves to 
the region a few GeV above the $t\bar{t}$ threshold to exemplify the size
of 
effects.  

Recall that the dominance of the $t \rightarrow bW^+$ decay mode leads to
a 
large top width $\Gamma_t$, which is about 1.5 GeV for a canonical mass
$m_t 
\simeq 175$ GeV.  This width is larger than the typical hadronic scale
$\mu \sim 
1$ fm$^{-1}$, and the top decays before it has time to hadronize [34],[35]
.  
It is precisely the large width that makes top physics so unique.
Firstly, the 
top decay width $\Gamma_t$ provides an infrared cut-off for the strong
forces 
between the $t$ and the $\bar{t}$ [36].    Secondly, $\Gamma_t$ controls
the
QCD  interferences between radiation occurring at different stages of the
$t\bar{t}$  production processes [37].  
These interferences affect the structure of the colour flows in the $t\bar
t$
events and may provide a potentially serious source of uncertainties in
the
reconstruction of the final state.

The interplay of several particle production sources is reminiscent of the 
 effects we have studied for process (3), but there are 
important differences.  From the onset, $W^+W^-$ events consist of two
separate 
colour singlets, $q_1\bar{q}_2$ and $q_3\bar{q}_4$, so that there is no
logical 
imperative of an interconnection between the two.  Something extra has to
happen
to  induce a colour rearrangement to $q_1\bar{q}_4$ and $q_3\bar{q}_2$
singlets, 
such as a perturbative exchange of gluons or a non-perturbative string
overlap.  
This introduces a sizeable dependence on the space-time picture, i.e.\ on
how 
far separated the $W^+$ and $W^-$ decay vertices are.  The process (11)
only 
involves one colour singlet.  Therefore the cross-talk is here inevitable.  
Recall also that, contrary to the  $W^+W^-$ case, there are no purely
leptonic
channels which could provide an interconnection-free environment.
Analogously to
the 
$W^+W^-$ case we expect that the perturbative restructuring is suppressed.  
However, a priori there is no obvious reason why interconnection effects
have to 
be small in the fragmentation process.  Moreover, the $b$ and $\bar{b}$
coming 
from the top decays carry compensating colour charges and therefore have
to \lq 
cross-talk' in order to produce a final state made up of colourless 
hadrons. 

Let us start from the perturbative picture.  In the process (11) the
standard 
parton showering can be generated by the systems of quarks appearing
within a 
short time scale, namely the $\widehat{t\bar{t}}, \, \widehat{tb}$ and 
$\widehat{\bar{t}\bar{b}}$ antennae/dipoles.

As was discussed in [37], the energy range of primary gluons, real or 
virtual, generated by the alternative quark systems of the type 
$\widehat{t\bar{b}}, \, \widehat{\bar{t}b}$ and $\widehat{b\bar{b}}$ is
strongly 
restricted, and one expects $\omega 
\lappeq \omega^{int}_{max} \sim \Gamma_t$.  Therefore the would-be parton 
showers initiated by such systems  and can hardly lead to a 
sizeable restructuring of the final state. In other words, the width of an
unstable particle acts as a kind of filter, which retains the bulk of the
radiation (with $\omega > \Gamma_t$) practically unaffected by the
relative
orientation of the daughter colour charges.

The general analysis of soft radiation in process (\ref{onze}) in terms of
QCD antennae was presented in [37]. Here we focus on the emission close to
the
$t\bar t$ threshold.

The primary-gluon radiation pattern can be presented as: 
\begin{equation} 
dN_g \equiv \frac{d\sigma_g}{\sigma_0} = \frac{d\omega}{\omega} 
\frac{d\Omega}{4\pi} \frac{C_F\alpha_s}{\pi} {\cal I} , 
\label{douze}
\end{equation} 
where $\Omega$ denotes the gluon solid angle; ${\cal I}$ is obtained by 
integrating the absolute square of the overall effective colour current
over the 
virtualities of the $t$ and $\bar{t}$. 

Near threshold   the $\widehat{tb}$ and $\widehat{\bar{t}\bar{b}}$
antennae are 
completely dominated by the emission off the $b$ quarks.  The distribution 
${\cal I}$ may then be presented in the form 
\begin{equation} 
{\cal I} = {\cal I}_{indep} + {\cal I}_{dec-dec} . 
\label{treize}
\end{equation} 
Here ${\cal I}_{indep}$ describes the case when the $b$ quarks radiate 
independently and ${\cal I}_{dec-dec}$ corresponds to the interference
between 
radiation accompanying the decay of the top and of the antitop 
\begin{equation} 
{\cal I}_{dec-dec} = 2 \chi (\omega) \frac{{\rm cos}\theta_1 {\rm
cos}\theta_2 - 
{\rm cos}\theta_{12}}{(1 - v_b{\rm cos}\theta_1)(1 - v_{\bar{b}} {\rm 
cos}\theta_2)} . 
\label{quatorze}
\end{equation} 
Here $\theta_1(\theta_2)$ is the angle between the $b(\bar{b})$ and the
gluon, 
$\theta_{12}$ is the angle between the $b$ and $\bar{b}$ and
$\chi(\omega)$ is 
the profile function [37], which controls the radiative interferences
between 
the different stages of process (\ref{onze}).

Near threshold 
\begin{equation} 
\chi(\omega) = \frac{\Gamma^2_t}{\Gamma^2_t + \omega^2} 
\label{quinze}
\end{equation} 

The profile function $\chi (\omega )$ cuts down the phase space available
for
emissions by the alternative quark systems and, thus, suppresses the
possibility
for such systems to develop QCD cascades. As $\Gamma_t\rightarrow\infty$,
the $b$
and $\bar b$ appear almost instantaneously, and they radiate coherently,
as
though produced directly. In particular, gluons from the $b$ and $\bar b$
interfere maximally, i.e., $\chi (\omega )$ = 1. At the other extreme, for 
$\Gamma_t\rightarrow 0$, the top has a long lifetime and the $b$ and $\bar
b$
appear in the course of the decays of top-flavoured hadrons at widely
separated
points in space and time. They therefore radiate independently. Thus a
finite top
width suppresses the interference compared to the na\"\i ve expectation of
fully
coherent emission. The same phenomena appear for the interference
contributions
corresponding to virtual diagrams.

The bulk of the radiation caused by primary gluons with $\omega >
\Gamma_t$ is 
governed by the $\widehat{tb}$ and $\widehat{\bar{t}\bar{b}}$ antennae.
It is 
thus practically unaffected by the relative orientation of the $b$ and
$\bar{b}$ 
jets.  In particular, the $\widehat{b\bar{b}}$ antenna is almost inactive.
The 
properties of individual $b$ jets are understood well enough, thanks to
our 
experience with $Z^0 \rightarrow b\bar{b}$ at LEP1. 

Because of the suppression of energetic emission associated with the
interferences, the restructuring could affect only soft particles.

Interconnection phenomena could affect the final state of $t\bar{t}$
events in 
many respects, but multiplicity distributions are especially transparent
to 
interpret.  As a specific example, we examined in Ref. [13] the total 
multiplicity of double leptonic top decays as a function of the relative
angle 
between the $b$ and $\bar{b}$ jets.  Let us make some comments concerning
the 
basic ideas of these studies: 
\begin{enumerate} 
\item As usual, one needs to model the fragmentation stage and study
quantities 
accessible at the hadron level.  
\item  A complication of attempting a full description is that it is no
longer 
enough to give the rate of primary-gluon emission: one must 
also allow for secondary branchings and specify the colour topology and 
fragmentation properties of radiated partons.  It is then useful to
benefit from 
the standard parton shower plus fragmentation picture for $e^+e^-
\rightarrow 
\gamma^*/Z^0 \rightarrow q\bar{q}$, where these aspects are understood.   
\item  The relation between $\gamma^*/Z^0 \rightarrow q\bar{q}$ and
$t\bar{t} 
\rightarrow bW^+\bar{b}W^-$ is most easily formulated in the
antenna/dipole 
language, see e.g., [38].  The independent emission term corresponds to
the sum of
two dipoles, 
${\cal I}_{indep} \propto \widehat{tb} + \widehat{\bar{t}\bar{b}}$, while
the 
decay-decay interference one corresponds to ${\cal I}_{dec-dec} \propto 
\chi(\omega)(\widehat{b\bar{b}} - \widehat{tb} -
\widehat{\bar{t}\bar{b}}$).  In 
total, therefore,  
\begin{equation} 
{\cal I} \propto (1 - \chi(\omega)) \widehat{tb} + (1 - \chi(\omega)) 
\widehat{\bar{t}\bar{b}} + \chi(\omega) \widehat{b\bar{b}} . 
\label{seize}
\end{equation} 
Each term here is positive definite and can be translated into a recipe
for 
parton shower evolution, see [13] for details. 
\item The top quarks are assumed to decay isotropically in their
respective rest 
frame, i.e., we do not attempt to include spin correlations between $t$
and
$\bar t$.   Breit-Wigner  distributions are included for the top and $W$
masses. 
\end{enumerate}

On the phenomenological side, the main conclusions of the analysis in [13]
are: 
\begin{itemize} 
\item The interconnection should be readily visible in the variation of
the 
average multiplicity as a function of the relative angle between the $b$
and 
$\bar{b}$.               
\item  A more detailed test is obtained by splitting the particle content
in 
momentum bins.  The high-momentum particles are mainly associated with the 
$\widehat{tb}$ and $\widehat{\bar{t}\bar{b}}$ dipoles and therefore follow
the 
$b$ and $\bar{b}$ directions, while the low-momentum ones are sensitive to
the 
assumed influence of the $\widehat{b\bar{b}}$ dipole. 
\item  A correct description of the event shapes in top decay, combined
with 
sensible reconstruction algorithms, could give errors on the top mass that
are 
on the level of 100 MeV (on top of possible BE effects). We recall here
also a
na\"\i ve perturbative estimate $(\delta m_t)_{PT} \sim
{\alpha_s(\Gamma_t)\over\pi} \Gamma_t\sim$ 70 MeV. 
\end{itemize} 

The possibility of interference reconnection effects in $t\bar{t}$
production is 
surely not restricted to the phenomena discussed here.  They could affect 
various other processes/characteristics.  

One topical example concerns the top quark momentum reconstruction.  As
was 
first emphasised in Ref.\ [33], the momentum measurement combined with the
threshold  scan could significantly improve the overall precision in
determination
of $m_t$  and  $\Gamma_t$.  As a supplementary bonus, the top momentum
proves to
be less sensitive to the beam effects.

In order to reconstruct the top momentum we need at least one of the
secondary 
$W$'s decaying hadronically.  
So the final state configurations are either a lepton plus four quark jets
or six
quark jets. In the latter case there is the $WW$ piece and the $b-W$ and
$b\bar
b$ interferences. Recall that the cross-talk between the $b$ and $\bar b$
jets is
not colour suppressed.
QCD interconnection may efface the separate identities of the top and
antitop 
systems and, thus, could produce a potential source of the systematic
error in 
the top momentum determination\footnote{The reconstruction of the
$W$-momentum 
could be affected as well.  The QCD interferences can have some impact
also on 
the top momentum distribution itself.}.  The interference pattern here is
more
complicated   than in the case of  double leptonic decays because of an
additional  cross-talking between the hadronically decaying $W$ and the
$b\bar{b}$ products. 

\vfill\eject

\noindent  {\bf 4~~~  Summary and Outlook}

The large width $(\Gamma \sim O$(1 GeV)) of the $W$ boson and of the top
quark
controls the radiative interferences between emission occuring at
different stages of the 
production processes.  The QCD interferences may efface the separate
identities 
of these particles and produce hadrons that cannot be uniquely assigned to 
either of them.  Here we concentrated mainly on two topical problems,
namely the 
QCD interconnection phenomena in events of the type $e^+e^- \rightarrow
W^+W^- 
\rightarrow q_1\bar{q}_2q_3\bar{q}_4$ and $e^+e^- \rightarrow t\bar{t} 
\rightarrow bW^+\bar{b}W^-$. 
On the perturbative level these interference effects are small, and one
has to
apply hadronization models to estimate the non-perturbative effects. 

The existing theoretical literature, based on quite different
philosophies, shows
a rather wide range of expectations for the shift in $m_W$, from a few MeV
to
several hundred MeV. The cross-talk between the $W$'s may have an 
impact on various other properties of hadronic $W^+W^-$ events as well.

Different hypotheses about the confinement dynamics may lead to different
expectations for the final-state event characteristics. So, in principle,
the
experimental tests of hadronic interconnection between the $W$'s --
connectometry
-- could provide a new laboratory for probing the structure of the QCD
vacuum.

In order to establish the evidence for a cross-talk in hadronic $W^+W^-$
events,
one has to find an observable which, on the one hand, proves to be quite
sensitive to this effect and, on the other hand, could allow rather
straightforward interpretation. The necessary requirements for such a
connectometer are that the no-reconnection predictions 
should be well understood, and that the expected signal is strong enough
to be
detectable within the limited statistics of LEP2. The latter is, by no
means, a
simple task.

In Ref. [24], an attempt was made to quantify the expectation based on the
string
hadronization model [25] in terms of the distributions of low-momentum
hadrons.
Essential advantages of such an approach to connectometry is that here the
no-reconnection case can be well described, and that there is no (direct)
dependence on the jet reconstruction method or event selection strategy.
The
first experimental results on connectometry in the $W^+W^-$ events have
already
been reported, see Refs. [3]-[6], and new experimental information
continues to
pour out from LEP2. The best we can hope is that the expected signal would
be at
the edge of observability. In such a case one would need a lot of hard
work (and
good luck) in order to detect the signal reliably. However, I would like
to
emphasize that, given the present lack of deep understanding of the
non-perturbative QCD dynamics, it is only experiment that could lead the
way.

It is anticipated that the systematic error on the top mass reconstruction
in the
process (\ref{onze}) would not exceed 100 MeV, see e.g. [13]. One may hope
that
with sophisticated analysis method such uncertainty can be reduced.

In some sense, the interconnection effects discussed here could be
considered as 
only the tip of the iceberg.  Colour reconnection can  occur in any
process 
which involves the simultaneous presence of more than one colour singlet.
Many 
of the techniques developed in Refs.~[12],[13],[22],[37] could be directly
applied
to  these problems.  

Among other examples of practical importance are $e^+e^- \rightarrow
Z^0H^0$, 
$e^+e^- \rightarrow Z^0Z^0$, $pp/\bar{p}p \rightarrow W^+W^-$,
$pp/\bar{p}p 
\rightarrow t\bar{t}$, $pp/\bar{p}p \rightarrow t\bar{b}$, $pp/p\bar{p} 
\rightarrow W^{\pm}H^0$, etc.  One could discuss also interferences with
beam 
jets.  The problem with these processes is that there are too many other 
uncertainties which make systematic studies look very difficult. 

Finally, let us recall that many aspects of the high accuracy
determination of
the parameters of the $W$ boson and of the top quark require a careful
analysis
of the QED radiative phenomena. Recall, for instance, that the $W$-width
effects seriously modify the  QED Coulomb corrections 
to the cross-section of the process $e^+e^- \rightarrow W^+W^-$, which
should be 
known with a high accuracy for the measurements scanning across the $WW$ 
threshold region [39].

The non-factorizable QED final-state interaction could induce some
systematic
effects in other $W$-mass measurements, for instance in $\bar pp$ collider
experiments. Of particular interest is the subprocess $qg\rightarrow W
q^\prime$
with $W\rightarrow \ell\nu_\ell$. Collider experiments normally rely on
the
equivalent process for $Z^0$ poroduction, $qg \rightarrow Zq$ with
$Z\rightarrow\ell^+\ell^-$, to calibrate the $W$ mass scale. Non-universal
interference effects are not included in such a procedure, e.g., a charged
$(W^\pm$) versus a chargeless ($Z^0$) intermediate state. Within current
experimental errors this would be negigible, but it could become relevant
for
future high-precision measurements.

\vspace*{1cm}
\noindent{\bf Acknowledgements}

I would like to thank A. De Angelis, J. Ellis, R. Orava and N. Watson for
useful
discussions. I am very grateful to T. Sj\"ostrand for fruitful
collaboration.

\vfill\eject

\end{document}